\documentclass[%
 reprint,
 amsmath,amssymb,
 aps,
]{revtex4-2}

\usepackage{graphicx}
\usepackage{dcolumn}
\usepackage{bm}
\usepackage{float}
\usepackage{xcolor}
\usepackage{ulem}

\newcommand{\add}[1]{\textcolor{black}{#1}}
\newcommand{\remove}[1]{\if0{#1}\fi}
\newcommand{\adds}[1]{\textcolor{black}{#1}}
\newcommand{\removes}[1]{\if0{#1}\fi}

\begin{document}


\title{Exactly solvable subspaces of non-integrable spin chains with boundaries and quasiparticle interactions}

\author{Chihiro Matsui}
\affiliation{%
 Graduate School of Mathematical Sciences, The University of Tokyo\\
 3-8-1, Komaba, Meguro-ku, 153-8914 Tokyo, Japan
}%




\date{\today}

\begin{abstract}
We propose two new strategies to construct a family of non-integrable spin chains with exactly solvable subspaces based on the idea of quasiparticle excitations from the matrix product vacuum state~\cite{bib:HOV13}. The first one allows the boundary generalization, while the second one makes it possible to construct the solvable subspace with interacting quasiparticles. Each generalization is realized by removing the assumption made in the conventional method~\cite{bib:MBBFR20}, which is the frustration-free condition or the local orthogonality, respectively. 
We found that the structure of the embedded equally-spaced energy spectrum is not violated by the diagonal boundaries, as long as quasiparticles are identical and non-interacting in the invariant subspace. On the other hand, we show that there exists a one-parameter family of non-integrable Hamiltonians which shows the perfectly embedded energy spectrum of the integrable spin chain. Surprisingly, the embedded energy spectrum does not change by varying the free parameter of the Hamiltonian. 
The constructed \add{models weakly break ergodicity,}
\removes{and therefore, the energy eigenstates in the solvable subspace are the candidates of non-thermal states in thermalizing systems,}
\adds{in which strong ETH is expected to be violated.} 
\end{abstract}

\maketitle


\section{Introduction}
Understanding the thermalization mechanism of isolated quantum systems is one of the most well-developed studies in recent statistical mechanics. After the eigenstate thermalization hypothesis (ETH) has been recasted as the most powerful candidate to explain thermalization phenomena, plenty of related works have been achieved including the ones which test validity or violation of the ETH. 
Although generic isolated quantum systems are believed to obey the strong ETH~\cite{bib:D91, bib:S94, bib:M09}, which requires that all the energy eigenstates are macroscopically indistinguishable from the thermal states, it has been found that some energy eigenstates are different from the thermal states by violating the statement of strong ETH. These non-thermal states often show up in the systems which do not thermalize, including the systems with integrability~\cite{bib:KWW06, bib:RDYO07} or many-body localization~\cite{bib:AABS19, bib:RDYO07, bib:BAA06, bib:NH15, bib:AV15, bib:SHBLFVASB15}, while it has been found that such non-thermal states also show up in the systems which do thermalize~\cite{bib:MRBR18, bib:MRB18, bib:IZ19, bib:ISX19, bib:OCMCCN19}. These non-thermal energy eigenstates are called {\it the quantum many-body scars}, named after the single-body quantum scar state~\cite{bib:H84}\add{, especially when they show long-lived oscillations for certain initial states~\cite{bib:TMASP18, bib:TMASP18-2}}. 

The first example of quantum many-body scars has been found experimentally for the Rydberg-atom quantum simulator~\cite{bib:BSKLOPCZEGVL17}, which shows the embedded equally-spaced energy spectrum. The system shows strong revivals and very slow thermalization when the initial state has non-negligible overlap with the eigenstates of the equally-spaced energies. This unforeseen behavior was expected to be caused by violation of ETH due to the \add{non-thermal property} of the energy eigenstates associated with the equally-spaced energies in the prepared initial state. 
Later, emergence of such non-thermal energy eigenstates has theoretically been explained by employing the $PXP$ model~\cite{bib:FSS04}, the effective model of the Rydberg atom chain, which admits exactly solvable energy eigenstates with equally-spaced energies~\cite{bib:TMASP18, bib:TMASP18-2}. 
Surprisingly, the known quantum many-body scars are often exactly solvable states of non-integrable systems. Besides the $PXP$ model, there exist a variety of models, including the AKLT model~\cite{bib:AKLT88, bib:MRBR18, bib:MRB18} and the Hubbard-type models~\cite{bib:VRB17, bib:YLB18}, which are non-integrable but have exactly solvable energy eigenstates. All those exactly solvable energy eigenstates are macroscopically distinguished from the thermal states. Therefore, we expect that exactly solvable states of non-integrable models are the candidates of \remove{quantum many-body scars }\add{non-thermal states in thermalizing systems}. 

It is believed that the thermalizing systems which admit emergence of \remove{quantum many-body scars }\add{non-thermal states} have the almost block-diagonal Hamiltonians~\cite{bib:SAP21, bib:MBR22}:  
\begin{equation} \label{eq:block_diagonal}
	\mathcal{H} \simeq W \oplus \mathcal{H}_{\rm thermal},  
\end{equation}
consisting of the large thermal subspace $\mathcal{H}_{\rm thermal}$ and the relatively small subspace $W$ which becomes negligible in the thermodynamic limit. Existence of the small invariant subspace $W$ weakly breaks the quantum version of ergodicity, as the states in this subspace cannot move out from $W$ during time evolution. Thus, the block diagonal Hamiltonian prevents full thermalization by keeping each energy eigenvector staying in each diagonal block. 
Recently, various methods to construct the Hamiltonian with the small invariant subspace have been proposed. The methods are mainly classified into three types, each of which is called the projector embedding~\cite{bib:SM17, bib:SGD20, bib:KMH20}, the spectrum generating algebra~\cite{bib:Y89, bib:A89, bib:Z90, bib:MRB18, bib:SI19, bib:IS20, bib:CPLH20, bib:SYK20, bib:CIKM23, bib:PZ22, bib:OM23}, or the Krylov restricted thermalization~\cite{bib:MPNRB21}. 
These are not always independent methods, but sometimes grasp different aspects of the same mathematical structure behind the Hamiltonians. Indeed, it can happen that a certain model is constructed by one method, and later, the same model is constructed by another method again. For instance, emergence of quantum many-body scars in the $PXP$ and AKLT model was first explained by the spectrum generating algebra~\cite{bib:MRBR18}, and then, the projector embedding type construction has been proposed for each model recently in \cite{bib:OM23, bib:OM23-2}. 

In this paper, we propose a new method to construct the Hamiltonian with the small invariant subspace based on the Bethe-ansatz method. The method is similar to the spectrum generating algebra, 
\begin{equation} \label{eq:spectral_g}
	\left([H,\,Q] - \mathcal{E}Q\right) \big|_W = 0, 
\end{equation}
in the sense that both methods provide the Hamiltonian and energy eigenstates in the subspace $W$ at the same time, although partial solvability of our method does not come from the spectrum generating algebra \eqref{eq:spectral_g}. 
The spectrum generating algebra also tells that the Hamiltonian has the equally-spaced energy spectrum in the solvable subspace $W$, which perfectly explains the strong revival obtained in the Rydberg atom experiment. The equally-spaced energy spectrum indicates that the quasiparticles living in the subspace $W$ are identical particles, while our method based on the Bethe ansatz provides the model with the solvable subspace in which the equally-spaced structure of the energy spectrum is broken, instead by showing the same energy spectrum as the spin-$1/2$ $XXX$ model. This implies that no revival phenomena will be obtained in the Bethe ansatz solvable subspace spanned by non-identical quasiparticle excitation states. 

\begin{table*}
\add{
\caption{Classification of partially solvable models with exact quasiparticle excitation states. The abbreviation SGA stands for the spectrum generating algebra. The models presented in this paper are highlighted by bold fonts. More detailed classification of the partially solvable models with non-interacting quasiparticles can be found in \cite{bib:CIKM23}. } 
\label{tab:psolvable_models}
\begin{tabular}{ccccc}
\hline
\vspace{1mm}
model & \parbox{30mm}{partial solvability mechanism} & \parbox{30mm}{quasiparticle interaction} & \parbox{30mm}{spectrum} & \parbox{30mm}{Matrix-product based expression} \\
\hline \hline
\vspace{2mm}
\bf{AKLT-type}~\cite{bib:MBBFR20} & \bf{SGA} & \bf{No} & \bf{equally-spaced} & \bf{Yes} \\
\vspace{2mm}
PXP model~\cite{bib:CIKM23, bib:PZ22, bib:OM23} & generalized SGA & No & equally-spaced & Yes \\
\vspace{2mm}
\parbox{30mm}{extensively many species~\cite{bib:ZM22}} & \parbox{30mm}{fully anti-symmetrized bases} & Yes & \parbox{30mm}{embedded $s=1/2$ Heisenberg} & No \\
\vspace{1mm}
\parbox{30mm}{\bf{$s=1$ nearest neighbor}} & \bf{Bethe ansatz} & \bf{Yes} & \parbox{30mm}{\bf{embedded $s=1/2$ Heisenberg}} & \bf{Yes} \\
\hline
\end{tabular}
}
\end{table*}

It should also be noted that most of the examples of \removes{non-thermal states of thermalizing systems }\adds{exactly solvable states for non-integrable systems} are written in the languages of non-interacting quasiparticles, while \removes{the candidates of non-thermal states }\adds{solvable states} constructed in this paper are expressed in terms of interacting quasiparticles. Only a few examples are known \removes{to be the non-thermal states }\adds{as exactly solvable states} consisting of interacting quasiparticles. One example is obtained in the deformation of the integrable Hamiltonian~\cite{bib:ZM22}, in which the exactly solvable energy eigenstates are constructed via the fully anti-symmetrized bases. 
Partial solvability of our model is completely independent from this example since its solvability comes from conventional integrability but its mathematical structure is highly non-trivial as we impose the integrability conditions on {\it the pseudo basis} constituted by the matrix-valued vectors. However, we would say that our model has advantage for practical uses, since the Hamiltonian simply consists of spin-$1$ nearest neighbor interactions. 
\removes{The solvable states have matrix product based expressions, which make it easier for us to compute their entanglement entropies.}
Besides, the method for constructing the partially solvable Hamiltonian presented in this paper can be applied to the other models associated with any integrable models.  \add{Differences among the partially models with solvable quasiparticle excitation states are summarized in Table \ref{tab:psolvable_models}}. 

This paper is organized as follows. In the next section, we define the model to be studied in this paper. We focus on the spin-$1$ chain which often shows up in the discussion of quantum many-body scars. The example includes the AKLT model. We also provide the basic notion of the matrix product state and quasiparticle excitation states, which are first introduced in the discussion of the generalized tangent space of the (nonlinear) manifold formed by the matrix product tensors~\cite{bib:HOV13, bib:MBBFR20}. 
\removes{We are especially interested in the small subspace spanned by these states, as they are expected to have relatively small entanglement entropies compared to the thermal states which have the volume-law entanglement entropies.}
\adds{Many of non-thermal states in thermalizing systems were found to be written in the matrix-product based expressions with fixed bond dimensions~\cite{bib:MBBFR20, bib:MBR22, bib:CIKM23}, which have relatively small entanglement entropies~\cite{bib:GVWC07, bib:VMC08, bib:S11, bib:O14, bib:MBR22} compared to those for the thermal states exhibiting the volume law behavior.}
\removes{For instance, a matrix product state is known to have the area-law entanglement entropy~\cite{bib:GVWC07, bib:VMC08, bib:S11, bib:O14}, if its bond dimension is finite, and quasiparticle excitation states are also expected to have the sub-volume-law entanglement entropies~\cite{bib:MBR22}. }
\adds{This fact strongly motivates us to look for the solvable states of non-integrable systems in the matrix product forms, as the candidates of non-thermal states in thermalizing systems. } 
\removes{Thus the matrix product state and quasiparticle excitation states are the candidates of non-thermal states since the small entanglement entropy is one of the characteristic features of non-thermal states. }
In Section \ref{sec:SGA}, we provide the Hamiltonian and its invariant subspace spanned by non-interacting quasiparticle excitation states. The first half of the section is devoted to the review of the known results for the periodic boundary models, whose partial solvability comes from the hidden spectrum generating algebra. In the last half, we discuss generalization to the non-trivial boundary cases. We show that the structure of the spectrum generating algebra is not violated by the diagonal boundary deformation. 
In Section \ref{sec:BAsolvable}, we discuss the construction of the Hamiltonian with the Bethe-ansatz solvable subspace. We show that the energy spectrum in the Bethe-ansatz solvable subspace coincides with the energy spectrum of the integrable system, without exhibiting the equally-spaced structure any more. The model which admits the Bethe-ansatz solvable subspace possesses a free-parameter, which does not show up in the energy spectrum of the solvable subspace. This implies that emergence of solvable subspace is robust against a certain kind of perturbations. We also remark that the energy spectrum in the Bethe-ansatz solvable subspace becomes continuous ranging to infinity in the thermodynamic limit, which is never obtained for the \removes{scar }\adds{solvable} subspace resulting from the spectrum generating algebra. 
The last section is devoted to the concluding remarks and future works.

\section{The model}
Let us consider the spin-$1$ chain with translationally invariant nearest neighbor bulk interactions. By writing the elementary matrix whose $(t,s)$-element is $1$ and the others are $0$ by $E^{t,s}$, the local bulk Hamiltonian is written as 
\begin{align} \label{eq:Hamiltonian}
	&h = \sum_{s,s',t,t'=0}^2 h^{s,s'}_{t,t'} E^{t,s} \otimes E^{t',s'}.    
\end{align} 
The whole Hamiltonian consists of the summation of the local Hamiltonian over all the sites. In this paper, we consider the periodic boundary: 
\begin{align}
	&H = \sum_{j=1}^{N} h_{j,j+1}
\end{align}
and the open boundaries: 
\begin{equation}
	H_{\rm B} = \sum_{j=1}^{N-1} h_{j,j+1}  + h_1 + h_N, 
\end{equation}
where $h_{j,j+1}$ is non-trivially acts on the $j$ and $j+1$th sites:
\begin{equation}
	h_{j,j+1} = \bm{1} \otimes \cdots \otimes \underset{j,j+1}{h} \otimes \cdots \otimes \bm{1}, 
\end{equation}
while $h_1$ and $h_N$ act on the $1$st and $N$th sites, respectively:
\begin{align}
	&h_1 = h_{\rm L} \otimes \bm{1} \otimes \cdots \otimes \bm{1}, \nonumber \\
	&h_N = \bm{1} \otimes \cdots \otimes \bm{1} \otimes h_{\rm R}. 
\end{align}
Besides the locality and translation invariance, we assume 
the spin-flip invariance: 
\begin{equation}
	h^{s,s'}_{t,t'} = h^{2-s,2-s'}_{2-t,2-t'}
\end{equation}
and conserved magnetization: 
\begin{equation} \label{eq:conserved_mag}
	h^{s,s'}_{t,t'} = h^{s,s'}_{t,t'} \delta_{s+s',t+t'}, 
\end{equation}
besides Hermiteness:  
\begin{equation} \label{eq:hermiteness}
	h^{s,s'}_{t,t'} = (h^{t,t'}_{s,s'})^* 
\end{equation}
for the local bulk Hamiltonian. These are natural assumptions realized by many models. 

Some of spin-$1$ chains equipped with the above properties are known to be integrable, including the Fateev-Zamolodchikov spin chain:  
\begin{align}
	h_{j,j+1} = \vec{S}_j \cdot \vec{S}_{j+1} - (\vec{S}_j \cdot \vec{S}_{j+1})^2, 
\end{align}
while some other spin-$1$ chains are known to have exactly solvable energy eigenstates, although they are non-integrable. One of the most famous examples of the latter case is the AKLT model: 
\begin{equation}
	 h_{j,j+1} = \vec{S}_j \cdot \vec{S}_{j+1} + \frac{1}{3} (\vec{S}_j \cdot \vec{S}_{j+1})^2, 
\end{equation}
which admits not only the exactly solvable ground state but also the exactly solvable excitation states associated with equally-spaced eigenenergies~\cite{bib:A89, bib:MRBR18, bib:MRB18, bib:MBBFR20, bib:MBR22}. 

Most of the known solvable energy eigenstates of non-integrable models are written in the homogeneous matrix product forms or quasiparticle excitations from the matrix product states~\cite{bib:HOV13, bib:MBBFR20}. 
The homogeneous matrix state is written in the following form: 
\begin{widetext}
\begin{align} \label{eq:vacuum_state}
	|\psi_A \rangle 
	&= {\rm tr}_a (K_a \vec{A} \otimes_p \cdots \otimes_p \vec{A}) \nonumber \\
	&= \sum_{(m_1,\dots,m_N) \in \{0,\dots,d-1\}^N} {\rm tr}_a (K_a A_{m_1} A_{m_2} \cdots A_{m_N})  |m_1,m_2,\dots,m_N \rangle, 
\end{align}
\end{widetext}
where \add{the $\chi$-by-$\chi$ matrices} $A_{m_n} \in {\rm End}(\mathbb{C}^{\chi})$ ($n=1,\dots,N$) act in the auxiliary space. Another index $d$ denotes the dimension of the local physical space. For the spin-$1$ chain, the local physical space must be three-dimensional, {\it i.e.} $d = 3$. Note that the trace ${\rm tr}_a$ is taken over the auxiliary space and the tensor product $\otimes_p$ must be operated on the physical spaces. 
The boundary matrix $K_a$, which acts in the auxiliary space $\mathbb{C}^{\chi}$, is determined by the boundary conditions. For instance, $K_a$ is the identity matrix for the periodic boundary, while $K_a$ is a certain matrix with ${\rm rank}\,K_a = 1$ for open boundaries. 
Throughout this paper, we focus on the matrix product states given by    
\begin{equation}
	\vec{A} = 
	\begin{pmatrix}
		a_0 \sigma^+ \\ a_1 \sigma^z \\ a_2 \sigma^-
	\end{pmatrix}, \qquad
	a_0,a_1,a_2 \in \mathbb{C},      
\end{equation}
which has the smallest non-trivial bond dimension $\chi = 2$. 
This class of the matrix product states includes the exactly solvable ground state of the AKLT model~\cite{bib:AKLT88}. 

On the other hand, we consider the one-quasiparticle excitation state expressed by 
\begin{align} \label{eq:quasip_ex}
	&| \psi_{A,B}(k) \rangle
	= \sum_{x=1}^N e^{ikx}\,  
	{\rm tr}_a (K_a \vec{A} \otimes_p \cdots \otimes_p \underset{x}{\vec{B}} \otimes_p \cdots \otimes_p \vec{A}),  
\end{align} 
where $\vec{B}$ is again the matrix-valued vector whose elements act in the auxiliary space $\mathbb{C}^{\chi}$, and locates at the position $x$ of the quasiparticle. In the above expression, no quasiparticle creation or annihilation is assumed, which is true for the periodic or diagonal boundaries. Indeed, the magnetization conservation property of the model, which we imposed in \eqref{eq:conserved_mag}, guarantees that the number of quasiparticles does not change in the bulk.   
The quasiparticle excitation state of the form \eqref{eq:quasip_ex} have been first proposed in the discussion of the generalized tangent space of the manifold formed by the matrix product tensors $\{ A_{m_1},\dots,A_{m_N} \}$~\cite{bib:HOV13}. 

The nature of quasiparticles depends on the choice of {\it the local quasiparticle creation operator} $O \in {\rm End}(\mathbb{C}^3)$ defined through the relation 
\begin{equation}
	\vec{B} = O \vec{A}.  
\end{equation}
For instance, quasiparticles show the non-interacting property under the nearest-neighbor Hamiltonian \eqref{eq:Hamiltonian} if the quasiparticle is chosen as the spin-$2$ magnon created by $O = (S^+)^2$~\cite{bib:MBBFR20}. The spin-$2$ magnon excitation states are known to form the solvable invariant subspace of the models belonging to the AKLT type~\cite{bib:MBBFR20}. Under the other choices of the creation operator, quasiparticles may interact with one another. The interacting quasiparticles are obtained in the Bethe-ansatz solvable subspace, as we will show in Section \ref{sec:BAsolvable}.

\section{\label{sec:SGA}Exactly solvable subspace without quasiparticle interaction}
In this section, we construct the solvable subspace $W$ spanned by non-interacting quasiparticle excitation states. 
The non-interacting property of quasiparticles is realized, for instance, by choosing the local quasiparticle operator $O$ as the spin-$2$ magnon creation operator: 
\begin{equation} \label{eq:spin-2magnon}
	O = (S^+)^2.   
\end{equation}
The other examples which produce non-interacting quasiparticles can be found in \cite{bib:MBBFR20}. The local spin-$2$ magnon creation operator satisfies the repulsive relations~\cite{bib:MBBFR20}: 
\begin{align}
	&O^2 \vec{A} = O \vec{B} = 0, \label{eq:repulsive1}\\
	&\vec{B} \otimes_p \vec{B} = 0, \label{eq:repulsive2}
\end{align}
which forbid quasiparticles to occupy the same site or adjacent sites. Thus, the spin-$2$ magnons do not interact each other since the Hamiltonian consists only of the nearest neighbor interactions \eqref{eq:Hamiltonian}. 

Throughout this section, we impose {\it the local orthogonality}: 
\begin{equation} \label{eq:local_orthogonality}
	({^t}\vec{A}^* \otimes_p {^t}\vec{A}^*) \cdot (\vec{B} \otimes_p \vec{A} + e^{ik} \vec{A} \otimes_p \vec{B}) = 0, 
\end{equation}
which is the sufficient condition for the quasiparticle excitation states \eqref{eq:quasip_ex} with the different number of quasiparticles to be orthogonal \add{since every inner product between the states with the different number of quasiparticles can be decomposed into the product of the local inner products including the left hand side of \eqref{eq:local_orthogonality} (see also Eq. (38) and Appendix C of \cite{bib:MBBFR20})}. Here the transpose in \eqref{eq:local_orthogonality} acts only in the auxiliary space.  
In the recent work of constructing a family of Hamiltonians with exactly solvable subspace~\cite{bib:MBBFR20}, the local orthogonality is always imposed. The local orthogonality allows only {\it identical} quasiparticles with momentum $k = \pi$ to exist \add{(see Eq. (52) of \cite{bib:MBBFR20})}. 
\add{This also means that there is a hidden spectrum generating algebra for this model}

With these properties, the multiple spin-$2$ magnon excitation states are represented as 
\begin{equation} \label{eq:multiquasip_ex}
	|\psi_{A,B^n} \rangle = Q^n |\psi_A \rangle,  
\end{equation}
in which the index $n$ represents the number of quasiparticles running over $n=1,\dots,\lfloor N/2 \rfloor$, due to the repulsive properties of quasiparticles \eqref{eq:repulsive1} and \eqref{eq:repulsive2}. 
$Q$ is the quasiparticle creation operator given by the summation of the local creation operator $O$ at each site: 
\begin{align} \label{eq:creation_op}
	&Q = \sum_{x=1}^N (-1)^x O_x,  \qquad
	O_x = \bm{1} \otimes \cdots \otimes \underset{x}{O} \otimes \cdots \otimes \bm{1},  
\end{align}
which is interpreted as the creation operator of the spin-$2$ magnon carrying the momentum $k = \pi$. 

\subsection{Periodic boundary case}
In this subsection, we discuss the periodic boundary case. The first part of this section is devoted to the review of the known models, which are the frustration-free models~\cite{bib:MBBFR20}. In the latter part of this section, we give the generalization of the known results by removing the frustration-free condition, which turns to be important for the boundary generalization, as we will see in the next subsection. 

In \cite{bib:MBBFR20}, it has been found that the sufficient conditions for the subspace 
\add{
\begin{equation} \label{eq:subspace_SGA}
	W_{\rm SGA} = {\rm span}\{ |\psi_{A} \rangle,\, Q|\psi_{A} \rangle,\, \dots, Q^{\lfloor N/2 \rfloor}|\psi_{A} \rangle \}
\end{equation}
}
to be the solvable subspace of the Hamiltonian are given by {\it the frustration-free condition}: 
\begin{equation} \label{eq:frustration_free}
	h \vec{A} \otimes_p \vec{A} = 0
\end{equation}
and {\it the eigenvalue condition}: 
\begin{equation} \label{eq:eigenvalue_cond}
	h (\vec{B} \otimes_p \vec{A} + e^{ik} \vec{A} \otimes_p \vec{B})
	= \mathcal{E} (\vec{B} \otimes_p \vec{A} + e^{ik} \vec{A} \otimes_p \vec{B}).  
\end{equation}
The first condition makes the vacuum state \eqref{eq:vacuum_state} be the zero-energy eigenstate, although it is not necessarily the ground state. The conditions \eqref{eq:frustration_free}, \eqref{eq:eigenvalue_cond} are equivalent to the spectrum generating algebra in the subspace \add{$W_{\rm SGA}$} \eqref{eq:subspace_SGA}: 
\begin{equation}
	\Big( [H,\,Q] - 2\mathcal{E} Q \Big) |\psi_{A,B^n} \rangle = 0,\quad
	n = 0,1,\dots,\left\lfloor \frac{N}{2} \right\rfloor. 
\end{equation}
Therefore, the energy spectrum of the Hamiltonian in \add{$W_{\rm SGA}$} shows the equally-spaced structure:  
\begin{equation}
	H |\psi_{A,B^n} \rangle = 2n\mathcal{E} |\psi_{A,B^n} \rangle, \quad
	n = 0, \dots, \left\lfloor \frac{N}{2} \right\rfloor,   
\end{equation}
which is understood also as the consequence of identical particle nature of spin-$2$ magnons. 
One thing which was missed to be noted in \cite{bib:MBBFR20} is that the quasiparticle excitation states \eqref{eq:quasip_ex} under the periodic boundary provides the energy eigenstate for the Hamiltonian only when the system consists of an even number $N$ of sites. 

The frustration-free condition and the eigenvalue condition are simultaneously solved by the local Hamiltonian
\adds{
\begin{align} \label{eq:frustration_freeH}
	h &= \frac{1}{2} h^{00}_{00} (S^x \otimes S^x + S^y \otimes S^y + S^z \otimes S^z) \\
		&- \left( \frac{1}{2} h^{00}_{00} + \frac{a_1^2}{a_0 a_2} h^{11}_{11} \right) (S^x \otimes S^x + S^y \otimes S^y + S^z \otimes S^z)^2 \nonumber  \\ 
		&- \left(  \frac{1}{2} h^{00}_{00} - \frac{a_1^2}{a_0 a_2} \Big(\frac{a_1^2}{a_0 a_2} - 1 \Big) h^{11}_{11} \right) (S^x \otimes S^x + S^y \otimes S^y)^2 \nonumber \\
		&- \left( h^{00}_{00} + \Big( \frac{a_1^2}{a_0 a_2} - 1 \Big) h^{11}_{11} \right) (S^z \otimes S^z)^2 \nonumber \\
		&+ \left( \frac{1}{2} h^{00}_{00} + \Big( \frac{a_1^4}{a_0^2 a_2^2} - 1 \Big) h^{11}_{11} \right) ((S^z)^2 \otimes \bm{1} + \bm{1} \otimes (S^z)^2) \nonumber \\
		&+ \left( 1 - 2 \frac{a_1^4}{a_0^2 a_2^2} \right) h^{11}_{11}\, \bm{1} \otimes \bm{1}, \nonumber
%
%
\end{align}
where $S^x$, $S^y$, and $S^z$ are the spin-$1$ operators associated with $\mathfrak{su}(2)$, while $\bm{1}$ represents the identity operator acting on a single physical space $\mathbb{C}^3$. }
The obtained local Hamiltonian \eqref{eq:frustration_freeH} contains essentially three free parameters, up to the overall factor, if one normalizes the quasiparticle excitation states \eqref{eq:quasip_ex}. This class of models includes the AKLT model, realized by choosing $h^{11}_{11} / h^{00}_{00} = 2/3$ and $a_0 = -\sqrt{2} a_1 = -a_2 = \sqrt{2/3}$, which perfectly explains the emergence of embedded equally-spaced energy spectrum obtained by the numerical test~\cite{bib:MRBR18}. 

Now the question is how much we can generalize a model in such a way that does not destroy the block diagonal structure \eqref{eq:block_diagonal}, {\it i.e.} that keeps \add{$W_{\rm SGA}$} as its invariant subspace. One possibility is to generalize the sufficient conditions \eqref{eq:frustration_free} and \eqref{eq:eigenvalue_cond} for \add{$W_{\rm SGA}$} to be the invariant subspace of the Hamiltonian. 
First, we replace the frustration-free condition with {\it the generalized frustration-free condition} 
\begin{equation} \label{eq:g-frustration_free}
	h \vec{A} \otimes_p \vec{A} = \vec{A} \otimes_p \vec{A'} - \vec{A'} \otimes_p \vec{A}.  
\end{equation}
Here $\vec{A'}$ is another matrix-valued vector whose elements are two-by-two matrices. 
This generalization \eqref{eq:g-frustration_free} reminds us the idea of constructing the steady states of the classical solvable stochastic processes such as the asymmetric simple exclusion process~\cite{bib:MGP68, bib:KDN90, bib:GS92, bib:DEHP93}. 
\add{Accordingly, modification of the eigenvalue condition \eqref{eq:eigenvalue_cond} as
\begin{equation} \label{eq:g-eigenvalue_cond}
	h (\vec{B} \otimes_p \vec{A} + e^{ik} \vec{A} \otimes_p \vec{B}) 
	=  \vec{B} \otimes_p \vec{Z} + e^{ik} \vec{X} \otimes_p \vec{B}   
\end{equation}
guarantees that the quasiparticle excitation states \eqref{eq:multiquasip_ex} to be the eigenstates of the Hamiltonian $H$, if the operator-valued vectors $\vec{X}$ and $\vec{Z}$ satisfy 
\begin{align}
	&\vec{Z} - \vec{A'} = \mathcal{E}'(k) \vec{A} \label{eq:bulk_cond1} \\
	&\vec{X} + \vec{A'} = \mathcal{E}(k) \vec{A}. \label{eq:bulk_cond2}
\end{align}
}
Besides these relations, we keep the local orthogonality \eqref{eq:local_orthogonality}, which allows only $k = \pi$ quasiparticles to exist. For this reason, we hereafter do not explicitly denote the dependence on $k$. 

The first condition \eqref{eq:g-frustration_free} again makes the vacuum state \eqref{eq:vacuum_state} be the zero-energy (but not necessarily the lowest energy) eigenstate under the periodic boundary condition. It also requires that the newly introduced matrix-valued vector $\vec{A'}$ to be 
\begin{equation}
	\vec{A'} = 
	\begin{pmatrix} b_0 \sigma+ \\ b_1 \sigma^z \\ b_2 \sigma^-  \end{pmatrix}, \quad
	b_0, b_1, b_2 \in \mathbb{C}, 
\end{equation}
where $b_2$ is restricted by the condition $b_0/a_0 = b_2/a_2$. 
The generalized frustration-free condition \eqref{eq:g-frustration_free}, together with the generalized eigenvalue condition \eqref{eq:g-eigenvalue_cond},  produces the hidden spectrum generating algebra:  
\begin{equation}
	\Big( [H,\,Q] - (\mathcal{E} + \mathcal{E}') Q \Big) |\psi_{A,B^n} \rangle = 0, \quad
	n=1,\dots,\left\lfloor  \frac{N}{2} \right\rfloor,  
\end{equation}
which implies that the embedded equally-spaced energy spectrum: 
\begin{equation}
	H |\psi_{A,B^n} \rangle = n (\mathcal{E} + \mathcal{E}') |\psi_{A,B^n} \rangle, \quad
	n = 1,\dots, \left\lfloor \frac{N}{2} \right\rfloor 
\end{equation}
is not violated by generalizing the frustration-free condition. 

The generalized conditions \eqref{eq:g-frustration_free} and \eqref{eq:g-eigenvalue_cond} are solve by the local Hamiltonian given by replacing the $(2,2)$ and $(8,8)$-elements of \eqref{eq:frustration_freeH} as $h^{00}_{00}/2 \to h^{00}_{00}/2 + {\rm Re}(b_0/a_0 - b_1/a_1)$, while the $(4,4)$ and $(6,6)$-elements as $h^{00}_{00}/2 \to h^{00}_{00}/2 - {\rm Re}(b_0/a_0 - b_1/a_1)$. 
Thus, the local bulk Hamiltonian under the generalized frustration-free condition contains two more free parameters besides the three parameters in the frustration-free case, if one fixes the normalization of the quasiparticle excitation states \eqref{eq:quasip_ex}. 
However, this increased freedom disappears under the presence of diagonal boundaries, when the four linearly independent vacua degenerate. We will see this point in the next subsection.

\subsection{Diagonal boundary case} \label{sec:diagonal_b}
When the open boundary condition is imposed, the boundary matrix in the matrix product state must be set as the rank $1$ matrix. Here we write the boundary matrix by: 
\begin{equation}
	K_a = |v_{\rm R} \rangle \langle v_{\rm L}|,  
\end{equation}
where the boundary vectors $| v_{\rm R} \rangle$ and $| v_{\rm L} \rangle$ are the vectors in the auxiliary space $\mathbb{C}^2$. 
Since the matrix product state takes different expressions depending on the choice of the boundaries, we explicitly denote the  boundary choice in the superscript: 
\begin{equation} \label{eq:MPS_open}
	|\psi_A^{(v_{\rm L}, v_{\rm R})} \rangle = {_a}\langle v_{\rm L}| \vec{A} \otimes_p \vec{A} \otimes_p \cdots \otimes_p \vec{A} |v_{\rm R} \rangle_a.    
\end{equation}

Throughout this subsection, we only consider diagonal boundaries: 
\begin{equation}
	h_{\rm L} =
	\begin{pmatrix} 
		\ell_0 & 0 & 0 \\
		0 & \ell_1 & 0 \\
		0 & 0 & \ell_2
	\end{pmatrix}, \quad
	h_{\rm R} =
	\begin{pmatrix} 
		r_0 & 0 & 0 \\
		0 & r_1 & 0 \\
		0 & 0 & r_2
	\end{pmatrix},   
\end{equation}
where $\ell_i, r_i \in \mathbb{R}$ ($i=0,1,2$).
Since the diagonal boundaries do not produce the quasiparticles, the expression for quasiparticle excitation state \eqref{eq:quasip_ex} is still valid. 

Now we look for the Hamiltonians which have the invariant subspace \add{$W_{\rm SGA}$} spanned by the matrix product state \eqref{eq:vacuum_state} and the quasiparticle excitations \eqref{eq:quasip_ex}. 
For the bulk solvability in the subspace \add{$W_{\rm SGA}$}, the generalized frustration-free condition \eqref{eq:g-frustration_free} and generalized eigenvalue condition \eqref{eq:g-eigenvalue_cond} must be satisfied. Besides, the boundary solvability requires the consistency conditions at the left boundary: 
\begin{align} 
	&{_a}\langle v_{\rm L}| (h_{\rm L} \vec{A} - \vec{A'}) = \mathcal{E}_{\rm L} \cdot {_a}\langle v_{\rm L}| \vec{A}, \nonumber \\
	&(h_{\rm R} \vec{A} + \vec{A'}) |v_{\rm R} \rangle_a = \mathcal{E}_{\rm R} \cdot \vec{A} |v_{\rm R} \rangle_a, \label{eq:boundary_cond1}
\end{align}
and the right boundary: 
\begin{align}
	&{_a}\langle v_{\rm L}| h_{\rm L} \vec{B} = (\mathcal{E} + \mathcal{E}_{\rm L})  \cdot {_a}\langle v_{\rm L}| \vec{B}, \nonumber \\
	&h_{\rm R} \vec{B} |v_{\rm R} \rangle_a = (\mathcal{E}' + \mathcal{E}_{\rm R}) \cdot \vec{B} |v_{\rm R} \rangle_a,  \label{eq:boundary_cond2}
\end{align}
respectively. 

The vacuum energy takes different values for the different choices of the boundary conditions. For instance, if we choose the diagonal boundaries which satisfy \eqref{eq:boundary_cond1} and \eqref{eq:boundary_cond2}, the vacuum energy is given by 
\begin{equation}
	H_{\rm B} |\psi_A^{(v_{\rm L},v_{\rm R})} \rangle = (\mathcal{E}_{\rm L} + \mathcal{E}_{\rm R}) |\psi_A^{(v_{\rm L},v_{\rm R})} \rangle. 
\end{equation}
For this reason, we call $\mathcal{E}_{\rm L}$ and $\mathcal{E}_{\rm R}$ the left and right boundary energies, respectively. 
From the boundary solvability conditions \eqref{eq:boundary_cond1} and \eqref{eq:boundary_cond2}, we find that the generalization of the frustration-free condition is important to obtain non-trivial boundary solutions, since the frustration free condition only allows the boundary interactions proportional to the identity matrix. 

The solutions to \eqref{eq:boundary_cond1} are classified into two types each for the left and right boundaries. The first type of solutions do not restrict the boundary vectors:  
\begin{align}
	&\mathcal{E}_{\rm L} = \ell_0 - \frac{b_0}{a_0} = \ell_1 - \frac{b_1}{a_1} = \ell_2 - \frac{b_2}{a_2}, \quad
	\forall\, |v_{\rm L} \rangle_a, \label{eq:left_g-cond} \\
	\text{resp.}\quad
	&\mathcal{E}_{\rm R} = r_0 + \frac{b_0}{a_0} = r_1 + \frac{b_1}{a_1} = r_2 + \frac{b_2}{a_2}, \quad
	\forall\, |v_{\rm R} \rangle_a, \label{eq:right_g-cond}
\end{align}
and thus, leads to the degenerate vacua with degree four. 
Indeed, the same degeneracy structure can be obtained in the ground state of the AKLT model under the presence of diagonal boundaries, since it is the special case of our model, as was mentioned in the previous subsection. 
The second type of the solutions determines the boundary vectors uniquely: 
\begin{align}
	&\mathcal{E}_{\rm L} = \ell_1 + \frac{b_1}{a_1} = \ell_2 + \frac{b_2}{a_2} \neq \ell_0 + \frac{b_0}{a_0}, \quad
	|v_{\rm L} \rangle_a = |1 \rangle_a
\end{align}
or 
\begin{equation}
	\mathcal{E}_{\rm L} = \ell_0 + \frac{b_0}{a_0} = \ell_1 + \frac{b_1}{a_1} \neq \ell_2 + \frac{b_2}{a_2}, \quad
	|v_{\rm L} \rangle_a = |0 \rangle_a, 
\end{equation}
{\it resp. }
\begin{align}
	&\mathcal{E}_{\rm R} = r_1 - \frac{b_1}{a_1} = r_2 + \frac{b_2}{a_2} \neq r_0 + \frac{b_0}{a_0}, \quad
	|v_{\rm R} \rangle_a = |0 \rangle_a 
\end{align}
or
\begin{equation}
	\mathcal{E}_{\rm R} = r_0 - \frac{b_0}{a_0} = r_1 + \frac{b_1}{a_1} \neq r_2 + \frac{b_2}{a_2}, \quad
	|v_{\rm R} \rangle_a = |1 \rangle_a, 
\end{equation}
and therefore, does not produce degeneracy for the vacuum states. 
In any case, we observe that the total boundary energy is determined by the elements of the boundary Hamiltonians as  
\begin{equation}
	\mathcal{E}_{\rm L} + \mathcal{E}_{\rm R} = \ell_1 + r_1.  
\end{equation}

In general, the degeneracy structure of the vacuum states does not survive for the quasiparticle excitation states. Only when we restrict the quasiparticle excitation energy as
\begin{align}
	\mathcal{E} + \mathcal{E}_{\rm L} = \ell_0, \qquad
	resp. \quad
	\mathcal{E}' + \mathcal{E}_{\rm R} = r_0,   
\end{align}
which is one of the solutions to \eqref{eq:boundary_cond2}, the quasiparticle excitation states with arbitrary boundary vectors can be the energy eigenstates, although the quasiparticles under this restriction carry zero energy $\mathcal{E} + \mathcal{E}' = 0$. 
The other solutions are given by 
\begin{align}
	&|v_{\rm L} \rangle_a = |0 \rangle_a, \qquad
	resp.\quad
	|v_{\rm R} \rangle_a = |1 \rangle_a, 
\end{align} 
for which not only the quasiparticle excitation states but also the vacuum states are not degenerate, as was obtained above. 

The hidden spectrum generating algebra in this model is produced by the bulk and boundary partial solvability conditions \eqref{eq:g-frustration_free}, \eqref{eq:g-eigenvalue_cond}, \eqref{eq:boundary_cond1}, and \eqref{eq:boundary_cond2}:  
\begin{equation}
	\Big( [H_{\rm B},\,Q] - (\mathcal{E} + \mathcal{E}') Q \Big) |\psi_{A,B^n}^{(v_{\rm L}, v_{\rm R})} \rangle = 0. 
\end{equation} 
Since the vacuum state $|\psi_A^{(v_{\rm L}, v_{\rm R})} \rangle$ has the eigenenergy given by $\mathcal{E}_{\rm L} + \mathcal{E}_{\rm R}$, the eigenenergies of the quasiparticle excitation states are obtained as 
\begin{equation}
	H_{\rm B} |\psi_{A,B^n}^{(v_{\rm L}, v_{\rm R})} \rangle = \Big(n (\mathcal{E} + \mathcal{E}') + \mathcal{E}_{\rm L} + \mathcal{E}_{\rm R} \Big) |\psi_{A,B^n}^{(v_{\rm L}, v_{\rm R})} \rangle. 
\end{equation}
Remind that the number of quasiparticles $n$ runs over $n = 1,\dots, \left\lfloor N/2 \right\rfloor$. 
In this way, the embedded equally-spaced energy spectrum structure is not violated by the diagonal boundaries. 

As was noted in the previous subsection, the boundary solvability conditions reduce the degrees of freedom for the desired Hamiltonian. For instance, when the four linearly independent vacua have the same energy, {\it i.e.} the conditions \eqref{eq:left_g-cond} and \eqref{eq:right_g-cond} are satisfied by the boundary Hamiltonians, the local Hamiltonian is just given by the frustration-free local Hamiltonian \eqref{eq:frustration_freeH} up to the constant $\ell_0 + r_0$ determined by choice of the boundaries. That is, the energy spectrum in the solvable subspace matches that for the frustration-free energy spectrum with the shift by $\ell_0 + r_0$.

\subsection{Off-diagonal boundary case}
Unlike the periodic or diagonal boundary cases, the off-diagonal boundaries create and annihilate quasiparticles. Therefore, the states with a fixed number of quasiparticles \eqref{eq:quasip_ex}, including the vacuum state \eqref{eq:vacuum_state}, are no more the eigenstates of the Hamiltonian. Instead, we assume the superposition of $n$-quasiparticle states as the energy eigenstate:  
\begin{align} \label{eq:offd_quasip_ex}
	&|\psi_{A,B^n}^{(v_{\rm L}, v_{\rm R})} \rangle = \sum_{n=0}^{\lfloor N/2 \rfloor} c_n Q^n\, {_a}\langle v_{\rm L}| \vec{A} \otimes_p \cdots \otimes_p \vec{A} |v_{\rm R} \rangle_a. 
\end{align}
The operator $Q$ is again the spin-$2$ magnon creation operator defined in \eqref{eq:multiquasip_ex}. We also impose the local orthogonality \eqref{eq:local_orthogonality}, which allows only $k = \pi$ identical quasiparticles to exist. 
We immediately notice that the superposition state \eqref{eq:offd_quasip_ex} becomes the energy eigenstate only when its bulk energy is zero: 
\begin{equation}
	\mathcal{E} + \mathcal{E'} = 0.   
\end{equation}
Besides, the left and right boundary solvability conditions: 
\begin{align}
	&{_a}\langle v_{\rm L}| (-c_n h_{\rm L} \vec{B} + c_{n-1} \vec{A'}) = \mathcal{E}_{\rm L}  \cdot {_a}\langle v_{\rm L}| (c_{n-1} \vec{A}), \\
	&{_a}\langle v_{\rm L}| (c_n h_{\rm L} \vec{A}) = \mathcal{E}_{\rm L} \cdot {_a}\langle v_{\rm L}|  (-c_{n+1} \vec{B}), 
\end{align}
and 
\begin{align}
	&((-1)^N c_n h_{\rm R} \vec{B} - c_{n-1} \vec{A'}) |v_{\rm R} \rangle_a = \mathcal{E}_{\rm R} \cdot (c_{n-1} \vec{A}) |v_{\rm R} \rangle_a, \\
	&c_n h_{\rm R} \vec{A} |v_{\rm R} \rangle_a = \mathcal{E}_{\rm R} \cdot ((-1)^N c_{n+1} \vec{B}) |v_{\rm R} \rangle_a, 
\end{align}
are required in order for \eqref{eq:offd_quasip_ex} to be the energy eigenstate. 

We found that the only non-trivial solutions to the bulk solvability condition \eqref{eq:g-frustration_free}, \eqref{eq:g-eigenvalue_cond} and boundary solvability conditions \eqref{eq:boundary_cond1}, \eqref{eq:boundary_cond2} are given by the choice of the boundary vectors $|v_{\rm L} \rangle = |1 \rangle$, $|v_{\rm R} \rangle = |0 \rangle$ and the boundary interactions: 
\begin{equation} \label{eq:b-int_off-d}
	h_{\rm L} = \begin{pmatrix} 0 & 0 & \ell_{02} \\ 0 & 0 & 0 & \\ \ell_{02}^* & 0 & 0 \end{pmatrix}, \qquad
	h_{\rm R} = \begin{pmatrix} 0 & 0 & r_{02} \\ 0 & 0 & 0 & \\ r_{02}^* & 0 & 0 \end{pmatrix} 
\end{equation}
under the restrictions on the boundary energies: 
\begin{equation}
	\mathcal{E}_{\rm L} = - \mathcal{E}_{\rm R} = \frac{b_1}{a_1}  
\end{equation}
and the ratios of the amplitudes: 
\begin{align}
	&\ell_{02} = -\mathcal{E}_{\rm L} \frac{c_{n+1}}{c_n}, \qquad
	r_{02} = (-1)^N\mathcal{E}_{\rm R} \frac{c_{n+1}}{c_n}, \\
	&\mathcal{E}_{\rm L} - \frac{\ell_{02} \ell_{02}^*}{\mathcal{E}_{\rm L}} \frac{c_{n}}{c^*_n} = - \mathcal{E}_{\rm R} + \frac{r_{02} r^*_{02}}{\mathcal{E}_{\rm R}} \frac{c_n}{c^*_n} = \frac{b_0}{a_0}. 
\end{align} 
That is, the only eigenvector consisting of the spin-$2$ magnon excitations is the zero-energy eigenstate: 
\begin{equation} 
	H_{\rm B}^{(1,0)} |\psi_{A,B^n} \rangle = 0. 
\end{equation}
Therefore, the solvable subspace of the off-diagonal boundary model is the one-dimensional space. Interestingly, the solvable state \eqref{eq:offd_quasip_ex} shows up in the middle of the energy spectrum under the generic choice of the off-diagonal boundary conditions (Appendix \ref{sec:offD_energy}). This implies that the superposition of quasiparticle excitations is again the candidate of a \remove{quantum many-body scar}\add{non-thermal state}.

\section{Exactly solvable subspace with quasiparticle interactions} \label{sec:BAsolvable}
In the previous section, we discussed the models with solvable subspace coming from the hidden spectrum generating algebra. The energy spectrum in the solvable subspace then shows the equally-spaced structure, and therefore, it is spanned by the identical quasiparticle excitation states with $k = \pi$. 
In this section, we propose new construction of the solvable subspace based on Bethe-ansatz solvability. The idea is to remove the local orthogonality \eqref{eq:local_orthogonality}, which is the sufficient condition for the quasiparticle excitation states with different number of quasiparticles to be orthogonal. Actually, the local orthogonality is too strong since the orthogonality of the energy eigenstates is guaranteed by the different eigenenergies, as we have chosen the Hermitian Hamiltonian \eqref{eq:hermiteness}. 
Instead, we impose the algebraic structure to produce integrability on the Hamiltonian in the subspace $W$. Then the matrix product state \eqref{eq:vacuum_state} and the quasiparticle excitation states: 
\begin{widetext}
\begin{align} \label{eq:bethe_state}
	&| \psi_{A,B^n}(\{k_\ell\}) \rangle
	= \sum_{1 \leq x_1 < x_2 < \cdots < x_n \leq N} f(x_1,x_2,\dots,x_n) \, 
	{\rm tr}_a(\vec{A} \otimes_p \cdots \otimes_p \underset{x_1}{\vec{B}} \otimes_p \cdots \otimes_p \underset{x_n}{\vec{B}} \otimes_p \cdots \otimes_p \vec{A}), 
	\\
	&f(x_1,x_2,\dots,x_n) 
	=  \sum_{P \in \mathfrak{S}_n} A_n(P)\, e^{i \sum_{j=1}^n k_{P(j)} x_j}, \nonumber
\end{align}
\end{widetext}
which are generalization of \eqref{eq:quasip_ex}, become the energy eigenstates in the subspace 
\add{
\begin{equation}
	W_{\rm BA} = {\rm span}\{ |\psi_{A} \rangle,\, |\psi_{A,B}(k_1) \rangle, \dots, |\psi_{A,B^{\lfloor N/2 \rfloor}}(\{k_\ell\}) \rangle \}, 
\end{equation}
}
if a set of quasiparticle momenta $\{k_j\}$ satisfies the Bethe equations. Here $\mathfrak{S}_n$ denotes the symmetric group of degree $n$. The amplitude $A_n(P)$ is determined by the boundary condition. For instance, the periodic boundary requires for $A_n(P)$ to satisfy 
\begin{equation}
	\frac{A_n(P \tau_{j,j+1})}{A_n(P)}
	=- \frac{1 + e^{i(k_{P(j)} + k_{P(j+1)})} - 2e^{ik_{P(j+1)}}}{1 + e^{i(k_{P(j)} + k_{P(j+1)})} - 2e^{ik_{P(j)}}}, 
\end{equation}
in which $\tau_{j,j+1}$ represents the transposition between the labels $j$ and $j+1$. 
The quasiparticle excitation states \eqref{eq:bethe_state} look similar to the Bethe states, but of course, they are not the Bethe states in the normal sense. 

The explicit forms of the Bethe equations depend on the models. Here we impose the spin-$1/2$ isotropic Heisenberg ($XXX$) like relations on the Hamiltonian in $W$: 
\begin{align}
	&h \vec{A} \otimes_p \vec{A} = 0 \label{eq:quad1} \\
	&h \vec{A} \otimes_p \vec{B} = -\vec{A} \otimes_p \vec{B} + \vec{B} \otimes_p \vec{A} \label{eq:quad2} \\
	&h \vec{B} \otimes_p \vec{A} = \vec{A} \otimes_p \vec{B} - \vec{B} \otimes_p \vec{A} \label{eq:quad3} \\
	&h \vec{B} \otimes_p \vec{B} = 0,  \label{eq:quad4}
\end{align}
although the Hamiltonian consists of $s=1$ spins. 
The first relation is nothing but the frustration-free condition imposed also in the previous subsection \eqref{eq:frustration_free}, and the last relation represents the repulsive property \eqref{eq:repulsive2} which forbid the quasiparticles to locate at the adjacent sites. 
The \add{spin-$1/2$} $XXX$-like relations \eqref{eq:quad1}-\eqref{eq:quad4} simultaneously determine the Hamiltonian and the local quasiparticle creation operator. 
The local quasiparticle creation operator which solves \eqref{eq:quad1}-\eqref{eq:quad4} is given by the diagonal matrix: 
\begin{equation}
	O  = 
	\begin{pmatrix} 
	\frac{b_0}{a_0} & 0 & 0 \\ 
	0 & \frac{b_1}{a_1} & 0 \\ 
	0 & 0 & \frac{b_0}{a_0} 
	\end{pmatrix},  
\end{equation}
which apparently allows double occupation for quasiparticles, since the repulsive relation \eqref{eq:repulsive1} does not hold for the above choice of $O$. This also indicates that the quasiparticles in the subspace \add{$W_{\rm BA}$} interact with each other. 
The Hamiltonian solves the relations \eqref{eq:quad1}-\eqref{eq:quad4} if the local Hamiltonian is given by 
\adds{
\begin{align} \label{eq:bethe_local_hamiltonian}
	h &= -( S^x \otimes S^x + S^y \otimes S^y) + \frac{1}{2} h^{00}_{00} S^z \otimes S^z \\
	&- (S^x \otimes S^x + S^y \otimes S^y + S^z \otimes S^z)^2 \nonumber \\
	&+ (S^x \otimes S^x + S^y \otimes S^y)^2 + \left( \frac{1}{2} h^{00}_{00} + 2 \right) (S^z \otimes S^z)^2 \nonumber \\
	&- ((S^z)^2 \otimes \bm{1} + \bm{1} \otimes (S^z)^2), \nonumber
%
\end{align}
}
which leaves the one parameter $h^{00}_{00}$ free. This local Hamiltonian is not in the class of the known integrable models~\cite{bib:U70, bib:S75, bib:FZ82}. 
\adds{Indeed, \eqref{eq:bethe_local_hamiltonian} satisfies the so-called {\it Reshetikhin condition}~\cite{bib:KS82, bib:GM95, bib:GP21}, known as the conjecture for Yang-Baxter solvability, only at $h^{00}_{00} = 0$. Actually, the Hamiltonian \eqref{eq:bethe_local_hamiltonian} becomes the integrable $XXC$ model~\cite{bib:M98} at this point~\cite{bib:PP23}. }

Then the periodic Hamiltonian consisting of \eqref{eq:bethe_local_hamiltonian} has \add{$W_{\rm BA}$} spanned by the Bethe-like states \eqref{eq:bethe_state} as its invariant subspace, if a set of quasiparticle momenta satisfies the Bethe equations for the spin-$1/2$ $XXX$ model: 
\begin{equation} \label{eq:Bethe_eq}
	e^{ik_jN} = -(-1)^n \prod_{\ell = 1 \atop \ell \neq j}^n \frac{e^{k_j + k_{\ell}} + 1 - 2e^{ik_j}}{e^{k_j + k_{\ell}} + 1 - 2e^{ik_{\ell}}}, \quad
	j = 1,\dots, n.     
\end{equation}
Of course, the energy spectrum of the Hamiltonian match the energy spectrum of the spin-$1/2$ $XXX$ model: 
\begin{equation}
	H |\psi_{A,B^n}(\{k_\ell\}) \rangle 
	= 4 \Big(\sum_{j=1}^n \cos k_j - n\Big) |\psi_{A,B^n}(\{k_\ell\}) \rangle 
\end{equation}
in the subspace \add{$W_{\rm BA}$}, which means the equally-spaced energy spectrum structure is broken in \add{$W_{\rm BA}$}. 
\add{This also implies that there is no hidden spectrum 
generating algebra for this model, }and therefore, \add{no revival phenomena will be obtained for this model}. 
The embedded spin-$1/2$ energy spectrum also indicates that the energy spectrum becomes gapless continuum ranging to infinity in the thermodynamic limit $N \to \infty$, as in the case of the spin-$1/2$ $XXX$ model, although the dimension of the subspace ${\rm dim}\,W < 2^N$ becomes negligibly small, compared to the dimension of its complement ${\rm dim}\,W^c > 3^N - 2^N$. \removes{This is very different structure from most of the known energy spectra for quantum many-body scars, which, by construction, often stay discrete even in the thermodynamic limit.} \adds{This is very different structure from the solvable models equipped with the hidden spectrum generating algebra, as they show the embedded equally-spaced energy spectra, which stay discrete in the thermodynamic limit. }

We have numerically checked the energy spectrum of the Hamiltonian \eqref{eq:bethe_local_hamiltonian} and indeed obtained the embedded energy spectrum of the spin-$1/2$ $XXX$ model in the full energy spectrum (Appendix \ref{sec:BA_energy}). 
Remarkably, the embedded $XXX$ energy spectrum is not violated by varying $h^{00}_{00}$, the free parameter in the local Hamiltonian. That is, this \remove{scar }\add{solvable} subspace is robust against the perturbation: 
\begin{equation}
	H_{\rm pert} = \alpha \sum_{j=1}^N (E_j^{0,0} E_{j+1}^{0,0} + E_j^{2,2} E_{j+1}^{2,2}), \quad
	\alpha \in \mathbb{R}. 
\end{equation}

It should be noted that the model \eqref{eq:bethe_local_hamiltonian} does not show the same degeneracy as that of the spin-$1/2$ $XXX$ model, since the subspace $W$ includes only the Bethe-like states, which correspond to the highest weight states of ${\mathfrak{sl}}_2$ for the $XXX$ model. Unfortunately, we are not succeeding to construct the operator for the model \eqref{eq:bethe_local_hamiltonian} which corresponds to $S_{\rm tot}^- = \sum_{x=1}^N S_x^-$ operator of the spin-$1/2$ $XXX$ model so far.

\section{Conclusion and discussion}
We have proposed the new construction of non-integrable spin chains with the exactly solvable subspace. The construction is based on the Bethe-ansatz method, which produces the invariant subspace not based on the spectrum generating algebra, and therefore, the energy spectrum in the solvable subspace is not equally spaced. 
As an example, we have constructed the spin-$1$ chain with the \add{spin-$1/2$} $XXX$-type solvable subspace, whose Hamiltonian shows the energy spectrum of the \add{spin-$1/2$} $XXX$ model embedded in the full energy spectrum. 
\removes{The subspace spanned by the known quantum many-body scars often shows the discrete energy spectrum consisting of a finite number of eigenenergies~\cite{bib:SM17} or infinitely many but equally-spaced eigenenergies~\cite{bib:MBBFR20}. 
On the other hand, }
\adds{The known partially solvable models equipped with the spectrum generating algebra in the solvable subspace show the discrete energy spectra consisting of infinitely many but equally-spaced eigenenergies~\cite{bib:MBBFR20}, while}
the model we have proposed in this paper shows the continuous energy spectrum even in the solvable subspace at the thermodynamic limit, as it coincides with the energy spectrum of the \add{spin-$1/2$} $XXX$ model, although the dimension of the subspace is negligibly small in the thermodynamic limit. This is the first difference which makes our model distinguished from the known non-integrable Hamiltonians with solvable subspaces. Subsequently, the broken hidden spectrum generating algebra in the subspace results in violation of the revival phenomena, which are often referred as the defining features of the models with quantum many-body scars~\cite{bib:TMASP18, bib:TMASP18-2}. \removes{Therefore, the model constructed in this paper provides a prototype of more general weakly ergodicity breaking systems.}
The second difference is obtained in the nature of quasiparticles in the solvable subspace. The known solvable subspace produced by the spectrum generating algebra is spanned by non-interacting quasiparticle excitation states, while the Bethe-ansatz solvable subspace we have constructed in this paper is spanned by interacting quasiparticle excitation states. 
These uncommon properties as the solvable subspace might be enough for saying that our model, \adds{which weakly violates ergodicity in the Hilbert space,} is a very new candidate of the thermalizing system with non-thermal energy eigenstates. 

We have also constructed the partially solvable spin chain with boundary magnetic fields. Partial solvability of this model comes from the hidden spectrum generating algebra, if the boundary Hamiltonians are diagonal. That is, the diagonal boundaries do not destroy the structure of the spectrum generating algebra in the model. Subsequently, the solvable subspace consists of identical and non-interacting quasiparticles, in which the Hamiltonian shows the equally-spaced energy spectrum. 
The situation is a bit different for the off-diagonal boundary case, since the solvable subspace of the off-diagonal boundary model is the one-dimensional space. However, the solvable state is in the middle of the spectrum, which can still be a candidate of \remove{a quantum many-body scar}\add{a non-thermal state}. 

Although we have provided the completely new construction of partially solvable models based on the algebraic structure of conventional integrable systems, \removes{we did not know whether the exactly solvable energy eigenstates found in this paper are non-thermal or thermal. Therefore, the first thing to be achieved in the next future work is to show that our exactly solvable energy eigenstates are non-thermal. }\adds{there are a number of interesting remaining problems to be addressed in the future. The first thing is to identify which solvable states among those constructed in this paper, are non-thermal.
We already observed several signs indicating that }the exactly solvable energy eigenstates constructed in this paper are the good candidates of non-thermal states. \removes{since }\adds{For instance,} the existence of infinitely long-lived quasiparticles is one of the characteristic features of non-thermal states~\cite{bib:CIKM23}. Especially when quasiparticles are interacting, their long life-time indicates that scattering processes are strictly restricted in such a way that quasiparticles scatter without decaying, which makes the states consisting of long-lived quasiparticles distinct from the typical (thermal) states. Besides, the exactly solvable energy eigenstates for the model with diagonal boundaries (Subsection \ref{sec:diagonal_b}) have entanglement entropies which obey the sub-volume-law, as was discussed in Subsection 3.3 of \cite{bib:MBR22}. This is another feature of non-thermal states, which is also expected for the Bethe-like states, as they have the matrix-product based expressions.  

\removes{The second thing we need to make sure is non-integrability of the model we have constructed in this paper. This is rather abstract problem which is difficult to be solved. Of course, we have checked that our model is not included in the class of the known integrable spin chains~\cite{bib:U70, bib:S75, bib:FZ82}, but this does not mean that the model is non-integrable, unless we remove all possibilities to find unknown integrable spin chains. One possible approach to this rather hard-looking problem is to check the energy level statistics, since energy level obeys different statistics, {\it i.e.} the Poisson distribution or Wigner-Dyson distribution, depending on whether the model is non-integrable or integrable, respectively. }\adds{Another future work is to check non-integrability of the models presented in this paper. We have already checked that the Reshetikhin condition holds for the proposed model only at the special point $h^{00}_{00} = 0$, indicating that the model is not Yang-Baxter solvable except for this point. Emergence of the Poisson distribution in the level-spacing statistics is another widely-used conjecture for testing chaotic nature of quantum systems~\cite{bib:BT77, bib:CCG85}, which is often associated with the non-integrability property. 
After all, no rigorous proofs are known for testing non-integrability of models, so far. }

From the mathematical point of view, it is a mystery where partial integrability of our model comes from. We have imposed the \add{spin-$1/2$} $XXX$-like relations on the matrix-valued vectors in the quasiparticle excitation states and accidentally found the solution, but of course, this does not mean we can always find the solution to the similar algebraic relations associated with the other integrable models such as the $XXZ$ model, supersymmetric $t$-$J$ model, Hubbard model, and so on. It would be nice to explain the existence of these solutions from the viewpoint of the Yang-Baxter equation, \removes{which is sometimes used as the definition of quantum integrability. }\adds{which enables us to use the methods developed for integrable systems to non-integrable models with embedded integrability. }

\begin{acknowledgments}
The author thanks C. Paletta and B. Pozsgay for helpful discussions about non-integrability of the obtained models. The author is also grateful to N. Tsuji for deriving the spin-operator expressions of the local Hamiltonians presented in this paper. 

C. M. is supported by JSPS KAKENHI Grant Number JP18K13465 and JP23K03244. 
\end{acknowledgments}

\appendix

\section{\label{sec:offD_energy}Energy spectrum of the $N = 4$ model with the spectrum generating algebra under the off-diagonal boundaries}
Table \ref{tab:off-diagonal} gives an example in which the solvable zero-energy eigenstate (highlighted by the bold fonts) shows up in the middle of the spectrum under the presence of the off-diagonal boundaries. The boundary Hamiltonians \eqref{eq:b-int_off-d} are chosen as $\ell_{02} = 3, r_{02} = -3$. The bulk parameters are chosen as $h^{00}_{00} = 5$ with $h^{11}_{11} / h^{00}_{00} = 2/3$, which is the AKLT point. Accordingly, the bulk Hamiltonian is set to satisfy the frustration-free condition, with the choice $a_0 = -\sqrt{2} a_1 = -a_2 = \sqrt{2/3}$. 

\begin{table}[h]
\caption{Energy spectrum of the model with the spectrum generating algebra under the off-diagonal boundaries.} 
\label{tab:off-diagonal}
\begin{ruledtabular}
\begin{tabular}{l}
16.849, 16.7337, 16.6279, 16.4139, 15.9501, 15.3909, 15., 15., \\
15., 14.3691, 14.3503, 14.2608, 14.2187, 14.0776, 13.5355, \\
12.9642, 12.9597, 12.7095, 12.349, 12.1843, 12.1705, 11.5636, \\
11.4777, 11.4735, 10.9842, 10.9122, 10.7428, 10.2352, 10., \\
9.97534, 9.67583, 9.66554, 9.5889, 9.4236, 9.35679, 9.25346, \\
8.89357, 8.88861, 8.64928, 8.41963, 8.03704, 7.87116, 7.86773, \\
7.84488, 7.63763, 7.26451, 7.23044, 6.97136, 6.7341, 6.63635, \\
6.57342, 6.54067, 6.46447, 6.16849, 6.15895, 6.00375, 5.72945, \\
5.45364, 5.38219, 5.21379, 5.21108, 5.01047, 5., 4.91893, \\
4.31464, 4.24784, 3.98426, 3.90293, 3.71403, 2.54418, -2.44381, \\
2.3304, 2.24423, 2.23303, 1.37816, -1.21692, 1.14052, 0.407693, \\
-0.0617388, 0.0616097, \bf{0.}
\end{tabular}
\end{ruledtabular}

\end{table}

\section{\label{sec:BA_energy}Energy spectrum of the $N = 5$ model with the Bethe-ansatz solvable subspace}
Table \ref{tab:BA1}-\ref{tab:BA3} give the energy spectra of the models with the Bethe-ansatz solvable subspace associated with the spin-$1/2$ $XXX$ model. The parameter $h^{00}_{00}$ in each table is chosen as $h^{00}_{00} = 0, 0.3,$ or $1.2$, respectively. The embedded spin-$1/2$ $XXX$ energy spectrum (highlighted by the bold fonts) is obtained, which is not violated by varying $h^{00}_{00}$. 

\begin{table*}[h]
\caption{Energy spectrum of the model with the Bethe-ansatz solvable subspace for $h^{00}_{00} = 0$.} 
\label{tab:BA1}
\add{
\begin{ruledtabular}
\begin{tabular}{l}
-13.6569, -13.6569, -13.6569, -13.6569, -13.6569, -13.6569, -13.6569, 
-13.1186, -13.1186, -13.1186, -13.1186, {\bf -12.4721}, -12.4721, \\
-11.617, -11.617, -11.617, -11.617, -9.5132, -9.5132, -9.5132, -9.5132, 
-9.5132, -9.5132, -9.5132, -9.5132, -9.5132, -9.5132, \\
-9.5132, -9.5132, -9.5132, -9.5132, {\bf -8.}, -8., -8., -8., -8., -8., -8., -8., 
-8., -7.95654, -7.95654, -7.95654, -7.95654, -7.80423, -7.80423, \\
-7.80423, -7.80423, -7.80423, -7.80423, -7.6405, -7.6405, -7.6405, 
-7.6405, -7.6405, -7.6405, -7.6405, -7.6405, -7.6405, -7.6405, \\
-7.6405, -7.6405, -7.6405, -7.6405, -7.63005, -7.63005, -7.63005, 
-7.63005, {\bf -7.23607}, -7.23607, -7.23607, -7.23607, -7.23607, \\
-7.23607, -7.23607, -7.23607, -7.23607, -7.23607, -7.23607, -7.23607, -7.23607, 
-7.23607, -6.35114, -6.35114, -6.35114, -6.35114, \\
-6.35114, -6.35114, -5.23607, -5.23607, -5.23607, -5.23607, -5.23607, -5.23607, -5.23607, 
-5.23607, -5.23607, -5.23607, -5.23607, \\
-5.23607, -5.23607, -5.23607, -5.23607, -5.23607, -4., -4., -4., -4., -4., -4., -3.95184, -3.95184, 
-3.95184, -3.95184, -3.72287, \\
-3.72287, -3.72287, -3.72287, -3.72287, -3.72287, -3.72287, -3.72287, -3.72287, -3.72287, -3.72287, -3.72287, 
-3.72287, -3.72287, \\
{\bf -3.52786}, -3.52786, -3.0595, -3.0595, -3.0595, -3.0595, {\bf -2.76393}, -2.76393, -2.76393, -2.76393, -2.76393, -2.76393, 
-2.76393, \\
-2.76393, -2.76393, -2.76393, -2.76393, -2.76393, -2.76393, -2.76393, -2.53562, -2.53562, -2.53562, -2.53562, -2.34315, -2.34315, \\
-2.34315, -2.34315, -2.34315, -2.34315, -2.34315, -1.64886, -1.64886, 
-1.64886, -1.64886, -1.64886, -1.64886, -1.12343, -1.12343, \\
-1.12343, -1.12343, -1.12343, -1.12343, -1.12343, -1.12343, -1.12343, -1.12343, 
-1.12343, -1.12343, -1.12343, -1.12343, -0.763932, \\
-0.763932, -0.763932, -0.763932, -0.763932, -0.763932, -0.763932, -0.763932, 
-0.763932, -0.763932, -0.763932, -0.763932, \\
-0.763932, -0.763932, -0.763932, -0.763932, -0.195774, -0.195774, -0.195774, -0.195774, 
-0.195774, -0.195774, -0.13087, -0.13087, \\
-0.13087, -0.13087, {\bf 0.}, 0., 
0., 0., 0., 0., 0., 0., 0., 0., 0., 0., 0., 0., 0., 0., 0., 0., 0., 
0., 0., 0., 0., 0., 0., 0., 0., 0., 0., 0., 0., 0., 0., \\
0., 0., 0., 0., 0.
\end{tabular}
\end{ruledtabular}
}

\medskip
\caption{Energy spectrum of the model with the Bethe-ansatz solvable subspace for $h^{00}_{00} = 0.3$.} 
\label{tab:BA2}
\add{
\begin{ruledtabular}
\begin{tabular}{l}
-13.6569, -13.5728, -13.5728, -13.4401, -13.4401, -12.9728, -12.9728, 
-12.9049, -12.9049, -12.9049, -12.9049, {\bf -12.4721}, -12.4721, \\
-11.4187, -11.4187, -11.4187, -11.4187, -9.5132, -9.5132, -9.48707, -9.48707, 
-9.48707, -9.48707, -9.31358, -9.31358, -9.31358, \\
-9.31358, -8.88707, -8.88707, -8.88707, -8.88707, {\bf -8.}, -8., -8., -8., -7.6405, -7.6405, 
-7.63463, -7.63463, -7.63463, -7.63463, \\
-7.53692, -7.53692, -7.53692, -7.53692, -7.38911, -7.38911, -7.38911, -7.38911, -7.24178, -7.24178, 
-7.24178, -7.24178, {\bf -7.23607}, \\
-7.23607, -7.23607, -7.23607, -7.12237, -7.12237, -7.11123, -7.08537, -7.08537, -6.93046, -6.93046, -6.93046, 
-6.93046, -6.91604, \\
-6.91604, -6.88195, -6.88195, -6.78911, -6.78911, -6.78911, -6.78911, -6.39211, -6.39211, -6.39211, -6.39211, -6.34994, 
-6.34994, \\
-6.2, -6.2, -5.4702, -5.4702, -5.41128, -5.41128, -5.41128, -5.41128, -5.23607, -5.23607, -5.23607, -5.23607, -5.23607, -5.23607, \\
-4.37195, -4.37195, -4.32783, -4.32783, -4.32783, -4.32783, -3.72287, 
-3.72287, -3.58347, -3.58347, -3.58347, -3.58347, {\bf -3.52786}, \\
-3.52786, -3.43607, -3.43607, -3.43607, -3.43607, -3.4, -3.38211, -3.38211, 
-3.38211, -3.38211, -3.149, -3.149, -3.149, -3.149, \\
-3.12237, -3.12237, -3.07763, -3.07763, 3., 3., -2.8, {\bf -2.76393}, -2.76393, 
-2.76393, -2.76393, -2.64315, -2.64315, -2.64315, -2.64315, \\
-2.62474, -2.62474, -2.549, -2.549, -2.549, -2.549, -2.34315, -2.31463, 
-2.31463, -2.15683, -2.15683, -2.15683, -2.15683, -1.87217, \\
-1.87217, -1.87217, -1.87217, -1.85326, -1.85326, -1.82805, -1.82805, -1.82718, 
-1.82718, 1.8, 1.8, 1.8, 1.8, 1.8, 1.8, 1.8, 1.8, 1.8, \\
1.8, 1.8, 1.8, 1.8, 1.8, 1.8, 1.8, 1.8, 1.8, 1.8, 1.8, -1.22718, -1.22718, -1.12343, 
-1.12343, 1.03607, 1.03607, 1.03607, 1.03607, \\
0.922375, 0.922375, 0.911234, -0.788721, -0.788721, -0.788721, -0.788721, -0.776122, 
-0.776122, -0.776122, -0.776122, -0.77482, \\
-0.77482, -0.77482, -0.77482, -0.763932, -0.763932, -0.763932, -0.763932, -0.763932, 
-0.763932, 0.730462, 0.730462, 0.730462, \\
0.730462, -0.729797, -0.729797, 0.716037, 0.716037, 0.6, 0.6, 0.6, 0.6, 0.6, 0.6, 0.6, 
0.6, 0.6, 0.6, 0.192214, 0.192214, 0.192214, \\
0.192214, 0.19211, 0.19211, 0.19211, 0.19211, -0.17482, -0.17482, -0.17482, -0.17482, 
0.149944, 0.149944, {\bf 0.}, 0., 0.
\end{tabular}
\end{ruledtabular}
}

\medskip
\caption{Energy spectrum of the model with the Bethe-ansatz solvable subspace for $h^{00}_{00} = 1.2$.} 
\label{tab:BA3}
\add{
\begin{ruledtabular}
\begin{tabular}{l}
-13.6569, -13.3605, -13.3605, -12.9235, -12.9235, {\bf -12.4721}, -12.4721, 
-12.4105, -12.4105, -12.4105, -12.4105, 12., 12., -11.0371, \\
-11.0371, -11.0371, -11.0371, -10.9605, -10.9605, -9.5132, -9.5132, -9.43117, 
-9.43117, -9.43117, -9.43117, -8.87214, -8.87214, \\
-8.87214, -8.87214, {\bf -8.}, -8., -8., -8., -7.6405, -7.6405, -7.32533, -7.32533, -7.32533, 
-7.32533, {\bf -7.23607}, -7.23607, -7.23607, \\
-7.23607, 7.2, 7.2, 7.2, 7.2, 7.2, 7.2, 7.2, 7.2, 7.2, 7.2, 7.2, 7.2, 7.2, 7.2, 7.2, 7.2, 7.2, 7.2, 
7.2, 7.2, -7.03117, -7.03117, \\
-7.03117, -7.03117, -6.78401, -6.78401, -6.78401, -6.78401, -6.77321, -6.77321, 6.43607, 6.43607, 6.43607, 
6.43607, -6.4061, \\
-6.4061, -6.4061, -6.4061, -5.96757, -5.96757, -5.96757, -5.96757, -5.74062, -5.74062, -5.23607, -5.23607, -5.23607, 
-5.23607, \\
-5.23607, -5.23607, -4.73238, -4.73238, -4.58304, -4.58304, -4.58304, -4.58304, -4.57612, -4.389, -4.389, -4.38401, -4.38401, \\
-4.38401, -4.38401, -4.25217, -4.25217, -4.25217, -4.25217, 3.93238, 
3.93238, -3.8514, -3.8514, 3.78304, 3.78304, 3.78304, 3.78304, \\
3.77612, -3.72287, -3.72287, 3.589, 3.589, {\bf -3.52786}, -3.52786, 
3.45217, 3.45217, 3.45217, 3.45217, 3.0514, 3.0514, -3.03967, \\
-3.03967, -2.85583, -2.85583, -2.85583, -2.85583, 2.82008, 2.82008, 
2.82008, 2.82008, -2.79016, -2.79016, -2.79016, -2.79016, \\
{\bf -2.76393}, -2.76393, -2.76393, -2.76393, -2.43671, -2.43671, -2.41017, -2.41017, 
-2.41017, -2.41017, 2.4, 2.4, 2.4, 2.4, 2.4, 2.4, 2.4, \\
2.4, 2.4, 2.4, -2.34315, 2.23967, 2.23967, 2.16049, 2.16049, -2.12275, -2.12275, 
1.96393, 1.96393, 1.96393, 1.96393, 1.61017, \\
1.61017, 1.61017, 1.61017, -1.6, -1.57566, -1.57566, -1.57566, -1.57566, -1.4049, 
-1.4049, -1.4049, -1.4049, 1.39331, 1.39331, \\
1.39331, 1.39331, 1.32275, 1.32275, -1.12343, -1.12343, -1.08782, -1.08782, -1.08782, 
-1.08782, 0.9951, 0.9951, 0.9951, 0.9951, \\
-0.828061, -0.828061, -0.828061, -0.828061, -0.826786, -0.826786, -0.8, 0.8, -0.8, 
0.775659, 0.775659, 0.775659, 0.775659, \\
-0.763932, -0.763932, -0.763932, -0.763932, -0.763932, -0.763932, -0.732381, -0.732381, 
0.587319, 0.587319, 0.587319, \\
0.587319, 0.420082, 0.420082, 0.420082, 0.420082, -0.239512, -0.239512, -0.0991273, -0.0991273, -0.0676192, -0.0676192, \\
{\bf 0.}, 0., 0. 
\end{tabular}
\end{ruledtabular}
}

\end{table*}

\newpage
\bibliography{references}

\end{document}